BAM 2023 CONFERENCE

US | BUSINESS SCHOOL
UNIVERSITY OF SUSSEX

Identity Dilemma in Immigrant consumers: a discourse analysis of diaspora marketing

**Mohammad Saleh Torkestani**
De Montfort University
**Zohreh Dehdashti Shahrokh**
Allameh Tabatabai University
**Kobra Bakhshi Zadeh**
Allameh Tabatabai University
**Pedram Jahedi**
Allameh Tabatabai University

# Identity Dilemma in Immigrant consumers: a discourse analysis of diaspora marketing

Mohammad Saleh Torkestani[1], Zohreh Dehdashti Shahrokh[2], Kobra Bakhshizadeh Borj[3], Pedram Jahedi[4]

**Abstract**

**Purpose:** Diasporas of a country-of-origin are groups interested in country-of-origin, scattered in host countries and familiar with markets of host countries. This paper aims to discourse analysis of diaspora marketing researches.
**Methodology:** In judgmental sampling using selected keywords, 24 researches in two eras before and after 2010 are selected for discourse analysis. Fairclough's critical discourse analysis model is used to analyze research data.
**Findings:** two different discourses can be distinguished in diaspora marketing researches. The diaspora niche marketing discourse is focused on penetrating into diaspora niche market. The diaspora marketing strategy discourse is focused on developing into international market using diaspora leverage.
**Research limitations and implications:** Judgmental sampling and lived experience of researchers in a developing country, which is usually considered as a diaspora's country-of-origin rather than a diaspora's host country limit the generalizability of research results.
**Practical implications:** Understanding the extensions of diaspora marketing concept and diaspora marketing discourses is helpful for designing appropriate strategies for entering and developing in the international market.
**Originality:** this paper investigates extensions of diaspora concept in previous research, identifies diaspora marketing discourses and compares dominant diaspora marketing discourse with global marketing discourse.

1. Lecturer in Marketing, School of Leadership, Management and Marketing, De Montfort University, Leicester, UK. (**Corresponding Author**) E-mail: mohammad.torkestani@dmu.ac.uk
2. Professor, Business management department, Faculty of Management and Accounting, Allameh Tabataba'i University, Tehran, Iran.
3. Associate Prof., Business management department, Faculty of Management and Accounting, Allameh Tabataba'i University, Tehran, Iran.
4. Ph.D. Student, Business management department, Faculty of Management and Accounting, Allameh Tabataba'i University. Tehran, Iran.

# 1. Introduction

Cultural diversity is one of the most important keywords to describe international market. Understanding the culture of the host country is also one of the most important determinants of success in international marketing. thus, it is appropriate to identify niches in the international market with cultural familiarity and similarity to the culture of country-of-origin. The diaspora of country-of-origin is a niche in the international market with the most cultural familiarity and similarity to country-of-origin (Risius, Hamm & Janssen, 2019). Diaspora is a group interested in country-of-origin and familiar with host country. Leaving the country-of-origin by diaspora does not mean a complete break from country-of-origin and complete attachment to the host country, but it means the journey of the culture of origin to adapt to the cultural context of the host country. For the country-of-origin, diaspora means populations with cultural and historical roots in a specific country-of-origin but scattered in several host countries. The cultural and historical roots of diaspora are the reason why their consumption preferences are similar to domestic market of country-of-origin. The similarity of diaspora consumption preferences is also a reason for considering the diaspora of each country-of-origin as a niche in international market. Diaspora marketing means the use of brands related to country-of-origin to leverage the cultural and historical roots of the diaspora in country-of-origin to attract their attention (Matejowsky, 2020). thus, the diaspora niche market is attractive for brands related to country-of-origin.

The concept of diaspora marketing in marketing research is often extension with a reductionist approach. In many related researches, the focus is on describing concepts such as nostalgia (Palomino-Tamayo, Saksanian & Regalado-Pezúa, 2021), country-of-origin (Matejowsky, 2020), etc. Focusing only on describing concepts related to diaspora marketing has caused researchers to pay less attention to diaspora marketing. Also, diaspora marketing is not easy in practice. On the one hand, diaspora marketing for country-of-origins' brands are one of the easiest ways to enter the international market. Because the diaspora niche market is similar with the consumer preferences of the domestic market of the country-of-origin. as well as, the diaspora has motivations such as nostalgia, identity, etc., towards brands related to country-of-origin (Elo et al., 2020). However, diaspora marketing often faces the difficulty of reaching the target market. diaspora niche market scattered across several host countries. Also, diaspora niche market is scattered in each of the host countries (Kroon, S., & Kurvers, 2020). That's why, diaspora niche market size is important. It should be considered that diaspora has the potential to influence consumption preferences of other citizens of host countries (Bowen, 2021). That's why, diaspora influence is important. There are different inferences about the expansion of the diaspora marketing concept. This research aims to discourse analysis of diaspora marketing researches.

## 2. Theoretical background

Every country is known with certain associations and its citizens are known with certain stereotypes. The associations of country affect the consumer's perception of the country's brands. The concept of brand's country-of-origin describes this affect. brand's country-of-origin is a concept to explain effect of associations related to a country and the stereotypes related to a country's citizens on perceptions toward brands related to that country (Risius, Hamm & Janssen, 2019). Country-of-origin refers to geographical affiliation of a brand in consumers' minds (Xu et al., 2020). Country-of-origin has no specific relevance with the place of production or the nationality of the brand owners (Hornikx et al., 2020). Country-of-origin is consumers' perception toward geographical origin of a brand (Thøgersen, Pedersen & Aschemann-Witzel, 2019). The stereotypes of the citizens of a country influence the marketers' perception toward consumer behavior of the citizens of that country. In international marketing, marketers' perception toward consumer behavior in a country has a reductionist approach (Seraphin, Korstanje & Gowreesunkar, 2020). Perhaps this reductionist approach is the reason for more focus on product performance and the emergence of the global marketing paradigm (Malhotra, Agarwal & Shainesh, 2018). However, international marketing consumers are not always citizens of other countries. A country-of-origins' Diaspora is a niche of the international market with the perception of belonging to country-of-origin and scattered in several host countries (Elo & Dana, 2019). Diaspora are all the dispersed populations in the host countries who have a sense about historical and cultural attribution to a country-of-origin (Grossman, 2019). Diaspora marketing is the lever of historical and cultural attribution of diaspora towards a country-of-origin by brands or products attributed to that country-of-origin. Diaspora marketing is an underrated concept in international marketing. While diaspora marketing concept is extension in many international marketing researches. The concept of diaspora marketing is widespread in at least three field of export market, tourism and investment. In the export market, more attention is paid to the spreading power of diaspora networks in international marketing. Successful diaspora marketing in export markets depends on understanding the mechanism of diaspora cooperation in foreign market entry and the impact of perceptions about diaspora culture in host countries (Elo el al., 2020). There are two processes for the development of diaspora tourism. The first process includes a simulated experience in host country, desire to return to country-of-origin, and experiences in country-of-origin. The second process is visit country-of-origin, experience in country-of-origin and simulated experiences in host country (Bowen, 2021). the priority of many developing countries is to attract diaspora investment. Diaspora is a vital factor in determining the policies of country-of-origin and host country towards each other. The country-of-origin is generally more willing to cooperate with diasporas' host countries. The government institutions of country-of-origin are more focused on identity-based motivations to attract diasporas' investment. while, the private organizations in country-of-origin are more focused on interests of diaspora (Poliakova, Riddle & Cummings, 2020). Also, diaspora marketing associate with at least four concepts of country-of-origin, diaspora, international marketing for brands and international marketing for countries. The success of the development programs of the country-of-origin depends on diaspora investment. while, diaspora investment depends on factors such as the stability of national currency, volume of trade and variety of production in country-of-origin. Good diplomatic relations between diasporas' country-of-origin and host country creates context for diaspora investment. while, Political development in country-of-origin with good state-nation relations Create context for diaspora Investment return (Ankomah et al., 2012). diasporas' needs are determined by structure of the diasporas' networks in host country. Factors such as the history of diasporas' migration (reasons, generations and etc.), acculturation and connection with country-of-origin are also influence diasporas' needs (Li, McKercher & Chan, 2020). Diaspora promotes country-of-origins' brands in country-of-origin and in host countries. country-of-origins' brands are nostalgic for diaspora. Diasporas' daily life in host countries and visit of country-of-origin is a constant invitation to empathize with the diaspora's nostalgia (Holak, 2014). In literature, the developing country is often the country-of-origin and the developed countries are the host country (Mullings, 2011). In developing countries, the retention of experts depends on the acceptance of the country's cultural values by the experts. While, the attraction of diasporas' experts depends on the acceptance of the country's cultural values by the diaspora. This is the reason for developing countries to focus on marketing their cultural values.

## 3. Methodology

3.1. Methodological approach

Methodological choice of this research is mono-method and focused on qualitative data gathering. Research aims to critical discourse analysis of diaspora marketing researches. Critical discourse analysis is the study of linguistic frameworks formed around a specific subject with a focus on dominant norms over time (Van Dijk, 2015). Critical discourse analysis is the study of what the language state is about a subject and why the language state is formed this way (Ivana & Suprayogi, 2020). Diaspora is citizen discourse in host country and country-of-origin which involves language-in-use and social contexts. Diaspora marketing is related to the concept of collective identity and power. Collective identity is the origin of similar consumption behavior among the diaspora. The concept of power exists in diaspora marketing researches between language-in-use and social contexts in host country and country-of-origin. Moreover, the concept of power is important in the affiliation of the diaspora marketing researchers to host country or country-of-origin.

3.2. Data collection

The statistical population of this research includes all diaspora marketing researches. The keywords diaspora, diaspora marketing, ethnic identity and ethnic marketing used to identify previous diaspora marketing researches in two eras before and after 2010. In our initial studies, a relative change in the researchers approach to diaspora marketing was identified in the era after 2010. Thus, two eras were defined for gathering previous diaspora marketing researches. judgmental sampling method used to identify previous diaspora marketing researches for critical discourse analysis. Finally, 10 researches before and 14 researches after 2010 are selected for critical discourse analysis of previous diaspora marketing researches. Table 1 presents selected research for critical discourse analysis.

Table 1. Researches included for critical discourse analysis

| Title | Author(s) | Year | Journal |
|---|---|---|---|
| Marketing the Diasporic Creed | Croucher & Haney | 1999 | Diaspora: A Journal of Transnational Studies |
| Africans in the diaspora: the diaspora and Africa | Akyeampong | 2000 | African affairs |
| The Cuban diaspora: A comparative analysis of the search for meaning among recent Cuban exiles and Cuban Americans | Bonnin & Brown | 2002 | Hispanic Journal of Behavioral Sciences |
| Marketing to the Welsh diaspora: The appeal to hiraeth and homecoming | Morgan, Pritchard & Pride | 2003 | Journal of Vacation Marketing |
| A new alliance for profit: China's local industries and the Chinese diaspora | Tracy & Lever-Tracy | 2003 | Chinese entrepreneurship and Asian business networks |
| Selling diaspora: Producing and segmenting the Jewish diaspora tourism market | Collins-Kreiner & Olsen | 2004 | Tourism, diasporas and space |
| Theorizing diaspora: perspectives on "classical" and "contemporary" diaspora | Reis | 2004 | International Migration |
| Building brand community among ethnic diaspora in the USA: Strategic implications for marketers | Quinn & Devasagayam | 2005 | Journal of Brand Management |
| Diaspora and development | Wei & Balasubramanyam | 2006 | World Economy |
| Diaspora and trade facilitation: The case of ethnic Chinese in Australia | Tung & Chung | 2010 | Asia Pacific Journal of Management |
| Diaspora entrepreneurs as institutional change agents: The case of Thamel. com | Riddle & Brinkerhoff | 2011 | International Business Review |
| Diaspora marketing | Kumar & Steenkamp | 2013 | Harvard Business Review |
| Diaspora tourists and the Scottish Homecoming 2009 | Sim & Leith | 2013 | Journal of Heritage Tourism |
| Diaspora institutions and diaspora governance | Gamlen | 2014 | International Migration Review |
| Typology of diaspora entrepreneurship: case studies in Uzbekistan | Elo | 2016 | Journal of International Entrepreneurship |
| Diaspora Marketing Revisited: The nexus of entrepreneurs and consumers | Sirkeci & Zeren | 2018 | Transnational Marketing Journal |
| Migration 'against the tide': Location and Jewish diaspora entrepreneurs | Elo, Täube & Volovelsky | 2019 | Regional Studies |
| Diaspora networks in international marketing: how do ethnic products diffuse to foreign markets? | Elo, Minto-Coy, Silva & Zhang | 2020 | European Journal of International Management |
| What's all the buzz about? Jollibee, diaspora marketing, and next-stage fast food globalization | Matejowsky | 2020 | Food and Foodways |
| Developing a multidimensional measurement scale for diaspora tourists' motivation | Otoo, Kim & Choi | 2021 | Journal of Travel Research |
| Food tourism: opportunities for SMEs through diaspora marketing? | Bowen | 2021 | British Food Journal |

| Title | Author(s) | Year | Journal |
|---|---|---|---|
| Psychological distance in the diaspora marketing of nostalgic products: a Venezuelan case | Palomino-Tamayo, Saksanian & Regalado-Pezúa | 2021 | Revista de Administração de Empresas |
| Who creates international marketing agility? Diasporic agility guiding new market entry processes in emerging contexts | Elo & Silva | 2022 | Thunderbird International Business Review |
| The Influence of Ethnic Identity on Consumer Behavior: Filipino and Lao Consumers in Australia | Intharacks, Chikweche & Stanton | 2022 | Journal of International Consumer Marketing |

3.3. Data analysis

The data analysis was conducted using Fairclough's critical discourse analysis model. This model includes three dimensions of description, interpretation and explanation. The analysis is done in each dimension separately. in description dimension should focus on the linguistic features of the text. In interpretation should focus on the processes of text production and consumption. finally, in explanation should focus on text's social practice (Fairclough, 1995). Texts are not completely understandable in isolation. Thus, analysis in three dimensions makes texts completely understandable. In this regard, for the description it should be focused on the concepts chosen by previous researchers to describe the diaspora marketing process. In interpretation, the first step is identifying usual concepts of each diaspora marketing discourses. It means extracting concepts related to the emergence of each discourse in the text. The second step is determining meanings related to usual concepts of each diaspora marketing discourses. In explanation, the first step is identifying the dominant discourse of diaspora marketing and international marketing. The second step is identifying usual marketing concepts associated with each of dominant discourses. The third step is Explaining the relevance of common marketing concepts of each dominant discourses.

3.4. Determining the quality of research

Research process in qualitative methods is less transparent in comparison with quantitative methods. Eriksson & Kovalainen (2015) argue that less transparency is the most important challenge in determining the quality of research in qualitative methods. Thus, the trustworthiness of qualitative research depends on the correct explanation of determining the quality of research. This issue is more acute in the critical discourse analysis. In this regard we gather enough data to ensure research theoretical contribution and credibility of the research findings. For dependability we used Fairclough's critical discourse analysis model in three dimensions of description, interpretation and explanation. We also took an integrated approach to the analysis of all texts (Fairclough, 1995). For confirmability used the supervision of an external auditor in all dimensions of critical discourse analysis and confirmation of research findings. Finally, for transferability we investigate similar diaspora marketing researches to compare with the findings of this research.

## 4. Findings

Two important points are understood in the initial review of the diaspora marketing research texts. New perspectives are emerging from around 2010 and diaspora marketing discourses can be described with frequent concepts, semantic inclusion, semantic contrast, anonymization, relational aspect and expressive aspect. studying texts related to eras before and after 2010 relatively shows extreme use of different concepts. Examining the frequent concepts of each eras shows redefinition of previous concepts in new concepts. After 2010, concepts are introduced such as Translation of product by diaspora for international market and product reconciliation according to diasporas' expectations in country-of-origins' domestic market (Riddle & Brinkerhoff, 2011., Sirkeci & Zeren, 2018., Bowen, 2021). While, other frequent concepts of this era are redefined according to historical context. In this redefining nationalism replaced by utilitarianism, historical identity replaced by cultural identity, acculturation replaced by cultural exchange, one country-of-origin replaced by several countries-of-origin, ethnic affirmers-assimilated replaced by bicultural-global, demographic replaced by behavioral, national replaced by global, penetration replaced by development, event replaced by Fast-moving consumer goods and consumer replaced by partner.

Examining the semantic inclusion of diaspora marketing researches before 2010 shows relative dominance of generalization of country-of-origin market with an emphasis on nationality, ethnicity, etc. Diaspora marketing research text in this era has semantic inclusion. The main concepts of generalization of country-of-origin market have a high coherence. Where the diaspora niche is described by nationalism (Tung & Chung, 2010), historical identity (Akyeampong, 2000), ethnic affirmers-assimilated (Bonnin & Brown, 2002), etc. These concepts describe diaspora niche market as a miniature of country-of-origin market. On the contrary, diaspora marketing researches after 2010 show the relative dominance of global marketing discourse. Also, the main concepts of generalization of country-of-origin market have a high coherence. In this era, diaspora has agency and is described by utilitarianism, belonging to several countries of origin (Elo, Täube & Volovelsky, 2019) and bicultural-global (Kumar & Steenkamp, 2013).

Examining diaspora marketing researches before 2010 shows a very low level of semantic contrast. In this era, only the contrast between acculturation with other main concepts is significant. Also, there is a significant semantic contrast between domestic marketing and international marketing. while, there is insignificant semantic contrast between the concepts related to internal marketing. After 2010, the semantic contrast is at a relatively lower level. In this era, utilitarianism contrasts with nationalism and Translation of product by diaspora for international market contrasts with product reconciliation according to diasporas' expectations in country-of-origins' domestic market. In this era, there is a low level of semantic contrast between diaspora marketing and international marketing. There is also a very low level of semantic contrast between diaspora marketing and global marketing discourse.

Examining diaspora marketing research before 2010 reflects a low level of anonymization. Therefore, the text provides a clear description of the effective use of national identity to penetrate in diaspora niche market. In this era, there is little ambiguity about the goals and outcomes of diaspora marketing. Meanwhile, in the era after 2010, anonymization is at a high level. In this era, there is no clear description of the goals and motivations of diaspora marketing. Therefore, there is a lot of ambiguity about the goals and outcomes of diaspora marketing. In this era, focusing on diaspora utilitarianism is the source of ambiguity about diaspora marketing goals. Also, focusing on cultural exchange is the source of ambiguity about diaspora marketing outcomes. In this era, anonymization is increasing in diaspora marketing researches.

Diaspora marketing research before 2010 reflects a relatively high level of relationship modality. In this era, the diaspora's demand is exclusive to brands of country-of-origin (Morgan, Pritchard & Pride, 2003). Also, the competition is exclusive to brands of country-of-origin. In fact, nationality and identity are the levers of power for brands of country-of-origin. Therefore, the agency of diaspora is negated. Also, diaspora marketing research after 2010 reflects a relatively high level of relationship modality. In this era, diaspora has utilitarian motives (Elo, Täube & Volovelsky, 2019). Also, the competition is not exclusive to brands of country-of-origin. Diaspora is the agent of translation of product for international market and idea creator for product reconciliation in country-of-origins' domestic market. Therefore, the focus is on the agency of diaspora in the domestic and international markets.

Diaspora marketing research before 2010 reflects a relatively high level of expression modality. In this era, there are dual filters to identify reality. For example, success of acculturation in host countries means emergence of assimilators and failure of acculturation in host countries means emergence of ethnic affirmers. Assimilators are interested in brands of host countries. While ethnic affirmers are interested in brands of country-of-origin. Diaspora marketing research after 2010 reflects a relatively average level of expression modality. In this era, outcomes of diaspora marketing are ambiguous, because diaspora is multicultural and global. Also, diaspora is a group with a number of countries-of-origin and agent of cultural exchange. In this era, success in international market depends on diasporas' participation. Thus, introduction of products to international market is diasporas' translation and reconciliation of product to domestic market is diasporas' expectations. Finally, main concepts of each era are listed in Table 2.

Table 2. Main marketing concepts related to each era

| Era | Concept |
|---|---|
| Before 2010 | Consumer, demographic, push, empathy, emotion-based value proposition, extension, country-of-origins' brand, market development, export, national brand, ethnic product, experiential, event, niche, international marketing. |
| After 2010 | Partner, behavioral, pull, participation, value proposition based on participation, adaptation, country-of-origins' brand, product development, diversity, global brand, nostalgia, cultural, motivational, influence marketing, word-of-mouth, interactive, network, global marketing |

The second dimension of interpretation consists of three steps. In the first step, diaspora marketing strategy discourse is identified as the dominant discourse of diaspora marketing. Diaspora marketing strategy discourse after 2010 is introducing diaspora marketing as a strategy to enter international market. In this discourse, Diaspora is an important partner for penetration and development in international market. The diaspora marketing strategy discourse is emerging against the diaspora niche marketing discourse. Before 2010, the diaspora niche marketing discourse introduces diaspora marketing as niche marketing in a familiar niche in international market. In this discourse, diaspora is only a consumer. In previous researches, global marketing discourse is introduced as the dominant discourse in international marketing. Global marketing discourse means trying to design a similar but flexible marketing mix to compete globally (Baena, 2019). In the second step, the main concepts of diaspora marketing strategy discourse are identified as the dominant discourse of diaspora marketing and the main concepts of global marketing are identified as the dominant discourse in international marketing. By studying previous researches, the main concepts of global consumer, global marketing mix, agile marketing and global competition are identified as main concepts of global marketing discourse (Elo, Harima & Freiling, 2015., Malhotra, Agarwal & Shainesh, 2018). While, the main concepts of bicultural-global consumer, marketing mix design according to cultural identity of consumer, translation of product by diaspora for international market and product reconciliation according to diasporas' expectations in country-of-origins' domestic market and competition in diaspora niche market for preparation of global competition are identified as main concepts of global marketing discourse. Table 4 shows main concepts of the dominant discourse of diaspora marketing and the dominant discourse of international marketing.

Table 3. Main concepts of the dominant discourses

| Discourse | Diaspora marketing strategy | Global marketing |
|---|---|---|
| Concept | Bicultural-global consumer | Global consumer |
| | Marketing mix design according to cultural identity of consumer | Global marketing mix |
| | Translation of product by diaspora for international market and product reconciliation according to diasporas' expectations in country-of-origins' domestic market | Agile marketing |
| | Competition in diaspora niche market for preparation of global competition | Global competition |

The third dimension is the explanation of diaspora marketing strategy discourse. thus, the main concepts of diaspora marketing strategy discourse are compared with the main concepts of global marketing discourse. The commonality between bicultural-global consumer in diaspora marketing strategy discourse and global consumer in global marketing discourse is utilitarian value in consumer behavior. Bicultural-global consumer in diaspora marketing strategy discourse does not negate the influence of cultural preferences on consumer behavior, but it shows the influence of different cultures on consumer behavior. marketing mix design according to cultural identity of consumer in diaspora marketing strategy discourse shows the important role of cultural exchange in value proposition, while the global marketing mix in global marketing discourse does not show this role. Translation of product by diaspora for international market and product reconciliation according to diasporas' expectations in country-of-origins' domestic market in diaspora marketing strategy discourse has limited the reference of learning to diaspora. While, this limitation is not in agile marketing. Finally, competition in diaspora niche market for preparation of global competition in diaspora marketing strategy discourse is the prelude to global competition in global marketing discourse.

## 5. Discussion and conclusion

5.1. Theoretical contributions

In this research, diaspora niche marketing discourse and diaspora marketing strategy discourse are identified. Discourse analysis of diaspora marketing research before 2010 reflects the relative dominance of diaspora niche marketing discourse. while, Discourse analysis of diaspora marketing research after 2010 reflects the relative dominance of diaspora marketing strategy discourse. The results reflect more alignment between diaspora marketing strategy discourse and global marketing discourse. This alignment is not limited to texts, but also reflected in metatextual and intertextual interpretations. thus, the emergence of diaspora marketing strategy discourse is a sign of the global marketing discourse dominance over diaspora marketing research. This dominance is a sign for diaspora marketing research linguistic frameworks more conformity with global marketing discourse in the future. It can be expected that more conformity with global marketing discourse will lead more marketing researchers to diaspora marketing research. Texts before 2010 are more focused on niches of international markets. in this era, diaspora marketing is mostly aimed to penetrating the international market and limited activity in the diaspora niche market (Tracy & Lever-Tracy, 2003., Quinn & Devasagayam, 2005). Concepts such as demographic segmentation (Akyeampong, 2000), assimilators-ethnic affirmers for targeting (Bonnin & Brown, 2002) and National brand for positioning (Morgan, Pritchard & Pride, 2003) are discussed in diaspora market researches. Also, concepts such as ethnic product (Quinn & Devasagayam, 2005), nationalism and identity (Tracy & Lever-Tracy, 2003), belonging to country-of-origin (Wei & Balasubramanyam, 2006) and diaspora businesses (Tung & Chung, 2010) are discussed. Texts after 2010 are more focused on development in international markets. In such a way that diaspora marketing is introduced as an attempt to enter the international market. In this era, diaspora marketing is a strategy to penetrate the diaspora market and leverage the diaspora to develop in international market (Sirkeci & Zeren 2018., Elo & Silva, 2022). Concepts such as diaspora networks and diaspora communication for segmentation (Palomino-Tamayo, Saksanian & Regalado-Pezúa, 2021), bicultural and global for targeting (Gamlen, 2014) and global brand for positioning (Elo, 2016) are discussed in diaspora market researches. Also, concepts such as global product and nostalgic product (Matejowsky, 2020., Bowen, 2021), utilitarianism (Riddle & Brinkerhoff, 2011), Cultural Heritage (Sim & Leith, 2013) and diaspora networks (Tung & Chung, 2010) are discussed.

## 5.2. Managerial implications

This research also has Practical tips for marketing managers. Diaspora niche market has features such as ease of achievement and difficulty of access. The ease of achievement potentially leads to Oversimplification of diaspora marketing and the difficulty of access potentially leads to neglect of diaspora marketing. Therefore, marketing managers should consider diaspora marketing discourses in the process of planning, implementation and evaluation. Diaspora marketing discourse selection should be based on internal and market evaluation. In this case, the descriptive words used in internal and market evaluation should be adapted to frequent concepts associated with each of diaspora marketing discourses. Also, it is more appropriate to follow a diaspora marketing discourse to shape and respond to expectations from diaspora marketing. Of course, following a diaspora marketing discourse is not only limited to words and text, but also involves aligning performance with discourse.

## 5.3. Limitations

Discourse analysis is a research method to the study of relationship between language-in-use and social contexts. language-in-use is relative and always changing. In fact, social contexts shape the appropriate language-in-use. Critical discourse analysis focuses on the norms of social contexts to shape the language-in-use. Therefore, the results of critical discourse analysis researches depend on researchers' lived experience. The results of this research depend on researchers' lived experience in a developing country. A country that is often considered as a country-of-origin rather than a host country. Judgmental sampling in this research uses the professional judgment of researchers to select the sample. But this judgment limits the generalizability of the research results. There is many research for the extended of diaspora marketing discourse as a strategy before 2010 and the extended of diaspora marketing discourse as a goal after 2010. Research like this are not a counterexample for the results of this research. Choosing the year 2010 as a basis for changing the discourse in the diaspora marketing researches is only to streamline the data collection and analysis process.

## 5.4. Future research

The dichotomy of development is very familiar to the scientific literature. Also, the developed and developing countries dichotomy is very familiar to marketing literature. This dichotomy is manifested in diaspora marketing researches in the form of host country and country-of-origin. This research is an attempt to analyze diaspora marketing from the perspective of researchers in a developing country or a country-of-origin. Future researches could analyze the influence of developed countries researchers' perspective on international marketing concepts. The diaspora market is an existing market for the country-of-origins' brands. While the international market is often a new market in the diaspora marketing researches. Thus, future researches should distinguish between the role of diaspora marketing strategy in international market penetration and international market development. Diaspora is a source of ideas for new products due to familiarity with other cultures. Diaspora is a bridge to access new markets due to connection with other countries. Thus, future researches should conduct case studies in the role of diaspora marketing in product development (in both host countries and country-of-origin) and diversification (in host countries).

**References**

Akyeampong, E. (2000). Africans in the diaspora: the diaspora and Africa. *African affairs*, *99*(395), 183-215.
Ankomah, P., Larson, T., Roberson, V., & Rotich, J. (2012). A creative approach to development: the case for active engagement of African diaspora in Ghana. *Journal of Black Studies*, *43*(4), 385-404.
Baena, V. (2019). Global marketing strategy in professional sports. Lessons from FC Bayern Munich. *Soccer & Society*, *20*(4), 660-674.
Bonnin, R., & Brown, C. (2002). The Cuban diaspora: A comparative analysis of the search for meaning among recent Cuban exiles and Cuban Americans. *Hispanic Journal of Behavioral Sciences*, *24*(4), 465-478.
Bowen, R. (2021). Food tourism: opportunities for SMEs through diaspora marketing?. *British Food Journal*. *124*(2), 514-529.
Collins-Kreiner, N., & Olsen, D. (2004). Selling diaspora: Producing and segmenting the Jewish diaspora tourism market. In *Tourism, diasporas and space* (pp. 293-304). Routledge.
Croucher, S. L., & Haney, P. J. (1999). Marketing the Diasporic Creed. *Diaspora: A Journal of Transnational Studies*, *8*(3), 309-330.
Elo, M. (2016). Typology of diaspora entrepreneurship: case studies in Uzbekistan. *Journal of International Entrepreneurship*, *14*(1), 121-155.
Elo, M., & Dana, L. P. (2019). Embeddedness and entrepreneurial traditions: Entrepreneurship of Bukharian Jews in diaspora. *Journal of Family Business Management*. , 22(1), 1-25.
Elo, M., Harima, A., & Freiling, J. (2015). To try or not to try? A story of diaspora entrepreneurship. In *The future of global organizing*, 10(1), 283-293.
Elo, M., Minto-Coy, I., Silva, S. C. E., & Zhang, X. (2020). Diaspora networks in international marketing: how do ethnic products diffuse to foreign markets?. *European Journal of International Management*, *14*(4), 693-729.
Elo, M., & Silva, S. (2022). Who creates international marketing agility? Diasporic agility guiding new market entry processes in emerging contexts. *Thunderbird International Business Review*, *12*(2), 1-12.
Elo, M., Täube, F., & Volovelsky, E. K. (2019). Migration 'against the tide': Location and Jewish diaspora entrepreneurs. *Regional Studies*, *53*(1), 95-106.
Eriksson, P., & Kovalainen, A. (2015). *Qualitative methods in business research: A practical guide to social research*. Sage.
Fairclough, N. (1995). (1995a) Critical Discourse Analysis. London: Longman.
Gamlen, A. (2014). Diaspora institutions and diaspora governance. *International Migration Review*, *48(1)*, 180-217.
Grossman, J. (2019). Toward a definition of diaspora. *Ethnic and Racial Studies*, *42*(8), 1263-1282.
Holak, S. L. (2014). From Brighton beach to blogs: Exploring food-related nostalgia in the Russian diaspora. *Consumption Markets & Culture*, *17*(2), 185-207.
Hornikx, J., van Meurs, F., van den Heuvel, J., & Janssen, A. (2020). How brands highlight country-of-origin in magazine advertising: A content analysis. Journal of Global Marketing, 33(1), 34-45.
Intharacks, J., Chikweche, T., & Stanton, J. (2022). The Influence of Ethnic Identity on Consumer Behavior: Filipino and Lao Consumers in Australia. *Journal of International Consumer Marketing*, 1-16.
Ivana, P. S. I., & Suprayogi, S. (2020). The Representation of Iran and United States in Donald Trump's Speech: A Critical Discourse Analysis. *Linguistics and Literature Journal*, *1*(2), 40-45.

Kroon, S., & Kurvers, J. (2020). Language use, language attitudes and identity in the East Timorese diaspora in the Netherlands. *Journal of Multilingual and Multicultural Development*, *41*(5), 444-456.

Kumar, N., & Steenkamp, J. B. E. (2013). Diaspora marketing. *Harvard Business Review*, *91*(10), 127.

Li, T. E., McKercher, B., & Chan, E. T. H. (2020). Towards a conceptual framework for diaspora tourism. *Current Issues in Tourism*, *23*(17), 2109-2126.

Malhotra, N. K., Agarwal, J., & Shainesh, G. (2018). Does country or culture matter in global marketing? An empirical investigation of service quality and satisfaction model with moderators in three countries. In *Emerging Issues in Global Marketing* (pp. 61-91). Springer, Cham.

Matejowsky, T. (2020). What's all the buzz about? Jollibee, diaspora marketing, and next-stage fast food globalization. *Food and Foodways*, *28*(4), 274-296.

Morgan, N., Pritchard, A., & Pride, R. (2003). Marketing to the Welsh diaspora: The appeal to hiraeth and homecoming. *Journal of Vacation Marketing*, *9*(1), 69-80.

Mullings, B. (2011). Diaspora strategies, skilled migrants and human capital enhancement in Jamaica. *Global Networks*, *11*(1), 24-42.

Otoo, F. E., Kim, S., & Choi, Y. (2021). Developing a multidimensional measurement scale for diaspora tourists' motivation. *Journal of Travel Research*, *60*(2), 417-433.

Palomino-Tamayo, W., Saksanian, M. C., & Regalado-Pezúa, O. (2021). Psychological distance in the diaspora marketing of nostalgic products: a Venezuelan case. *Revista de Administração de Empresas*, *62*(1), 1-19.

Poliakova, E., Riddle, L., & Cummings, M. E. (2020). Diaspora investment promotion via public–private partnerships: Case-study insights and IB research implications from the Succeed in Ireland initiative. *Journal of International Business Policy*, *3*(1), 23-37.

Quinn, M., & Devasagayam, R. (2005). Building brand community among ethnic diaspora in the USA: Strategic implications for marketers. *Journal of Brand Management*, 13(2), 101-114.

Reis, M. (2004). Theorizing diaspora: perspectives on "classical" and "contemporary" diaspora. *International Migration*, *42*(2), 41-60.

Riddle, L., & Brinkerhoff, J. (2011). Diaspora entrepreneurs as institutional change agents: The case of Thamel. com. *International Business Review*, *20*(6), 670-680.

Risius, A., Hamm, U., & Janssen, M. (2019). Target groups for fish from aquaculture: Consumer segmentation based on sustainability attributes and country-of-origin. *Aquaculture*, *499*, 341-347.

Seraphin, H., Korstanje, M., & Gowreesunkar, V. (2020). Diaspora and ambidextrous management of tourism in post-colonial, post-conflict and post-disaster destinations. *Journal of Tourism and Cultural Change*, *18*(2), 113-132.

Sim, D., & Leith, M. (2013). Diaspora tourists and the Scottish Homecoming 2009. *Journal of Heritage Tourism*, *8*(4), 259-274.

Sirkeci, I., & Zeren, F. (2018). Diaspora Marketing Revisited: The nexus of entrepreneurs and consumers. *Transnational Marketing Journal*, *6*(2), 139-157.

Thøgersen, J., Pedersen, S., & Aschemann-Witzel, J. (2019). The impact of organic certification and country-of-origin on consumer food choice in developed and emerging economies. *Food Quality and Preference*, *72*, 10-30.

Tracy, N., & Lever-Tracy, C. (2003). A new alliance for profit: China's local industries and the Chinese diaspora. In *Chinese entrepreneurship and Asian business networks* (pp. 81-99). Routledge.

Tung, R. L., & Chung, H. F. (2010). Diaspora and trade facilitation: The case of ethnic Chinese in Australia. *Asia Pacific Journal of Management*, *27*(3), 371-392.

Van Dijk, T. A. (2015). Critical discourse analysis. *The handbook of discourse analysis*, 466-485.

Wei, Y., & Balasubramanyam, V. N. (2006). Diaspora and development. *World Economy*, *29*(11), 1599-1609.

Xu, X., Comello, M. L. G., Lee, S., & Clancy, R. (2020). Exploring country-of-origin perceptions and ethnocentrism: The case of US dairy marketing in China. *Journal of Food Products Marketing*, *26*(2), 79-102.